\documentstyle[11pt,paspconf,epsf]{article}

\def\etal{{\it et al.}}
\def\rms{{\it rms\/}}

\begin{document}

\title{The Evolution of Stellar Populations in Intermediate Redshift Clusters}

\author{Daniel D. Kelson}
\affil{Dept. of Terrestrial Magnetism, Carnegie Institution of
Washington, 5241 Broad Branch Rd., NW, Washington, DC 20015}

\author{Garth D. Illingworth}
\affil{UCO/Lick Observatory, Dept. of Astronomy and
Astrophysics, UCSC, Santa Cruz, CA 95064}

\author{Marijn Franx and Pieter G. van Dokkum}
\affil{Leiden Observatory, P.O. Box 9513, NL-2300 RA,
Leiden, The Netherlands}

\begin{abstract} The $M/L$ ratios and absorption line-strengths of
distant cluster galaxies can be used to directly study their stellar
populations, determine their redshift of formation, their scatter in
ages, and any dependence of their ages on such internal properties
such as velocity dispersion or mass. Comparing the zero-point the
fundamental plane in the $z=0.33$ cluster CL1358+62 to that in Coma,
we conclude that the redshift of formation for the stars in massive
cluster E/S0s was $z\ga 2$. The fundamental plane in CL1358+62 has the
following form: $r_e \propto \sigma^{1.31\pm 0.13}\langle
I\rangle_e^{-0.86\pm 0.10}$, indicating that the last epoch of
star-formation has very little dependence on galaxy mass. The scatter
in the CL1358+62 fundamental plane is also very low, equivalent to a
scatter in ages of $\sim 15\%$.We have also analyzed the $M/L_V$
ratios of galaxies of type S0/a and later. These early-type spirals
follow a different plane from the E and S0 galaxies, with a scatter
that is twice as large as the scatter for the E/S0s. These residuals
also correlate with the residuals from the color-magnitude relation.
Preliminary analysis of the absorption line-strengths of the CL1358+62
early-types indicate that they are well-described by uniformly old,
single-burst stellar populations with metallicity varying as a
function of velocity dispersion.\end{abstract}



\section{Introduction}

The fundamental plane (FP) is an empirical relation between galaxy
half-light radius, $r_e$, surface brightness, $\langle I\rangle_e$,
and central velocity dispersion, $\sigma$ for early-type galaxies.
Locally, $$ r_e\propto \sigma^{1.24} \langle I\rangle_e^{-0.82} $$ in
Gunn $r_g$ (J\o{}rgensen \etal\ 1996). Under the assumption of
homology, this implies $ M/L_{r_g} \propto M^{1/4} r_e^{-0.02}$. The
fundamental plane is very thin, with an \rms\ scatter of $\pm 23\%$ in
Coma in $M/L_V$ ratio for a given galaxy mass, $M$ (J\o{}rgensen
\etal\ 1993). With its low scatter, the FP is a very powerful tool for
measuring the evolution of galaxy $M/L$ ratios as a function of
redshift, $M$, etc.

Because reasonable stellar populations evolve as $M/L_V \propto
t^\kappa$ (Tinsley \& Gunn 1976), evolution in the properties of the
fundamental plane directly probe the star-formation histories of
cluster galaxies. Measurements of the evolution of the FP zero-point,
scatter, and slope can be used to constrain the mean epoch of
star-formation, the spread in early-type galaxy ages; and the
dependence of galaxy age on, {\rm e.g.\/}, $M$.

We exploit high resolution Keck spectroscopy for velocity dispersions
and line strengths, and WFPC2 imaging for the structural parameters,
colors, and morphologies. Details can be found in Kelson \etal\
(1999abc).


\section{The Fundamental Plane of Early-Type Galaxies in CL1358+62}

\centerline{
\vbox{
\vbox{\textwidth=2in \plotone{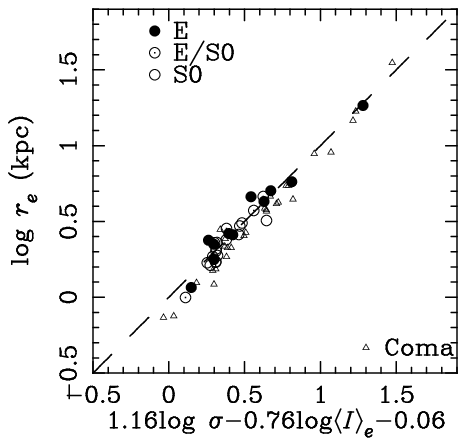} \quad \plotone{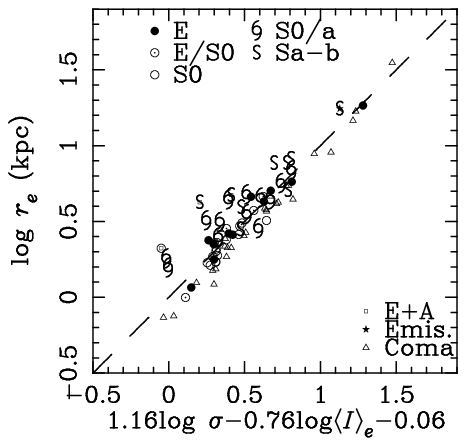} }
\vbox{\footnotesize
\noindent Figure 1.\ \ \
The left-hand panel shows the fundamental plane of E/S0s in CL1358+62.
The early-type galaxies clearly form a tight relation. The intrinsic
scatter is 14\% in $r_e$. In the right-hand panel, the full sample of
53 galaxies is shown, including the early-type spirals.}}
}
\medskip

The left-hand panel of Figure 1 shows the tight fundamental plane
relation in the $z=0.33$ cluster CL1358+62. The 30 E/S0s in this
sample are fit by the following plane in Johnson $V$ (Kelson \etal\
1999c): $$r_e \propto \sigma^{1.31\pm 0.13}\langle
I\rangle_e^{-0.86\pm 0.10}$$ This sample has comparable depth to the
FP samples of Coma, ensuring that the comparison of these FP exponents
with those determined in nearby clusters is reasonably free from
selection effects. We draw the following three conclusions:
\smallskip

\noindent
$\bullet$ Mild evolution in the FP zero-point, of $\Delta M/L_V =
-0.13 \pm 0.03$ ($q_0=0.1$), indicates a mean redshift of formation
of $\langle z_f\rangle \ga 2$ for the stars in cluster E/S0s.
\smallskip

\noindent
$\bullet$ The scatter in the color-magnitude relation and FP suggest
a 1-$\sigma$ scatter in ages of 15\%. For $q_0=0.05$ and $\langle
z_f\rangle =2$, $\pm 1$-$\sigma$ is equivalent to $1.5 \la z_f\la
3.5$.
\smallskip

\noindent
$\bullet$ No evolution in the FP slope is seen, indicating that the
mean luminosity weighted ages of the stellar populations do not depend
significantly on $\sigma$ or $M$:
$$ \log M/L\big|_{z=0.33} -
\log M/L\big|_{z=0} \propto
 (0.01\pm 0.23)\log \sigma - (0.16\pm 0.16)\log r_e$$


\section{The Spiral Galaxies of CL1358+62}

Our sample contains many galaxies of type S0/a through Sb. These
early-type spirals are shown with the E/S0s in the right-hand panel of
Figure 1. The spirals show large scatter about the FP and have
systematically higher surface brightnesses for a given $\sigma$ and
$r_e$. The scatter for the early-type spirals is twice as large as
that for the E/S0s. For the spirals, the FP residuals correlate
strongly with residual from the color-magnitude relation (see Kelson
\etal\ 1999c).

Correcting the surface brightnesses of the spirals for their young
stellar populations using their residuals from the color-magnitude
relation, one obtains a new plane for the spirals which follows that
of the E/S0s. We conclude that after several Gyr, these spirals,
including the low-mass E+As, may evolve into present-day, low-mass S0s
and appear in nearby FP samples.


\section{Absorption Line Strengths of Early-Type Galaxies in
CL1358+62}

The CL1358+62 spectra are of sufficiently high $S/N$ that one can
accurately measure absorption line strengths, and use stellar
population synthesis models, such as those of Vazdekis \etal\ (1996),
to constrain the systematic variation of the stellar populations along
the FP and color-magnitude relations.

\medskip
\centerline{
\vbox{
\hbox{\valign{#\vfil\cr
\hbox{ \textwidth=1.9in \plotone{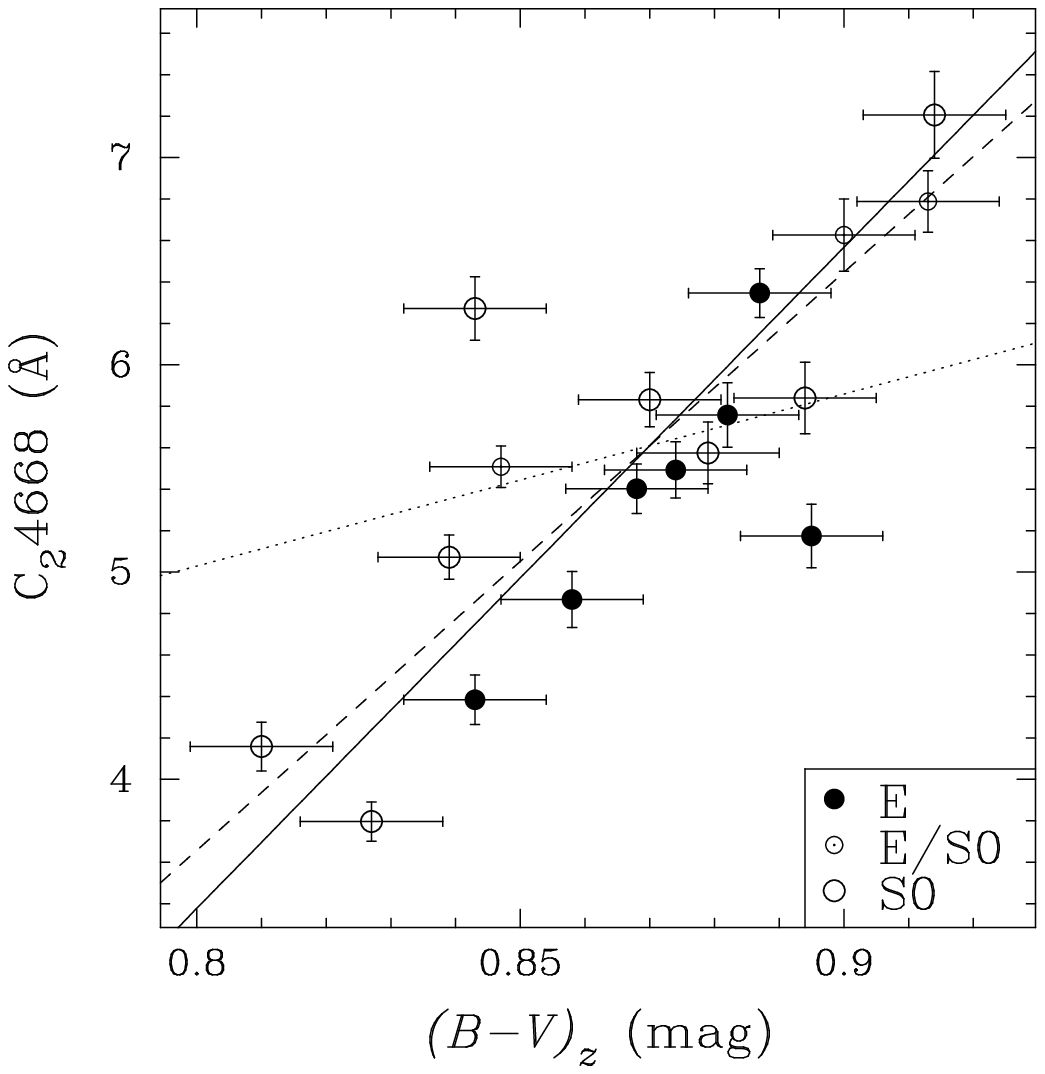} }\cr
\hbox{\qquad}\cr
\vbox{\hsize=3in\footnotesize
\noindent Figure 2.\ \ \
Measurements of the Lick/IDS C$_2$4668\AA\ index are
plotted against restframe $(B-V)$ color for the E/S0 galaxies. The
Vazdekis \etal\ (1996) single-burst stellar population models predict
the slope shown by the dashed line if metallicity is the sole cause
for variations in these colors and indices. The dotted line shows that
expected for age alone. The best-fit slope in our preliminary
analysis, shown by the solid line, is nearly identical to the slope
defined by metallicity variations at fixed age (Vazdekis \etal\
1996).}\cr
}}
} }

In Figure 2, we show that the sequence of E/S0 galaxies in CL1358+62
is defined by a tight correlation between C$_2$4668 \AA\ and restframe
$(B-V)$ color. The best-fit slope of the correlation is nearly
identical to that predicted for a sequence of single-burst stellar
population models of constant age, in which only metallicity varies.
Even with these preliminary measurements, this conclusion is
reinforced by the correlations of the metal-sensitive C$_2$4668\AA\
index with the high-order Balmer lines. In the first panel of Figure
3, the C$_2$4668\AA\ index is shown to systematically correlate with
the H$\delta_A$ index in a manner which is also consistent with the
line of constant age. The scatter in these two diagrams is consistent
with the small scatter in ages that has been estimated using the FP
and color-magnitude relations (Kelson \etal\ 1999c, van Dokkum \etal\
1998).

The latter two panels show the correlations of C$_2$4668\AA\ and
H$\delta_A$ with central velocity dispersion, indicating that
metallicity is strongly correlated with the depth of the galaxy
potential. The solid lines are the best-fit slopes, with the $\pm
1$-$\sigma$ uncertainties indicated by dotted lines. Using the
Vazdekis \etal\ (1996) models, these correlations with velocity
dispersion imply $M/L_V \propto M^{0.14\pm 0.02}$. Preliminary
analysis using Monte Carlo simulations of real data (Kelson 1999)
indicate that this correlation can produce an observed FP of the
approximate form: $r_e\propto \sigma^{1.3}\langle I\rangle_e^{-0.8}$,
though the precise shape depends on sample selection criteria, and the
underlying distribution of bulge-to-disk ratios.

\medskip
\centerline{
\vbox{
\vbox{\textwidth=5in \plotone{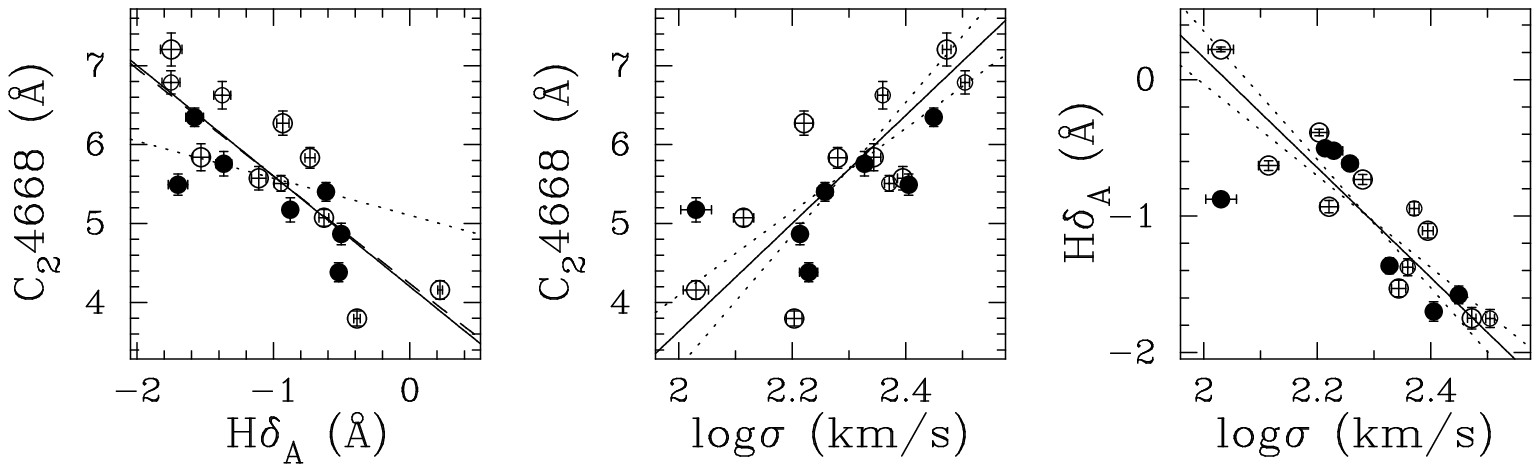} }
\vbox{\footnotesize
\noindent Figure 3.\ \ \
Using the E/S0 galaxies we show, in the left-most panel, the observed
correlations of the metal-sensitive C$_2$4668 \AA\ Lick/IDS index with
the H$\delta_A$ index of Worthey \& Ottavianni (1997). The two
right-hand panels show these two indices plotted against central
velocity dispersion for the same galaxies.
}}}
\smallskip

\section{Summary}

We have found that the slope of the fundamental plane has remained
constant with redshift. Together with the absorption line strengths
and restframe colors, these data suggest that early-type galaxies in
clusters have very uniform stellar population ages, and chemical
abundances which are strongly determined by the depths of their
potential wells, assuming that the stellar populations of cluster
E/S0s can be adequately modeled by simple single-burst models. The
implied correlation of metallicity with $\sigma$ appears to be the
physical basis behind both the color-magnitude relation and the
fundamental plane, though further work is required to fully understand
the nature of the selection biases and other systematic effects.



\begin{references}
\reference J\o{}rgensen I., Franx M., \& Kj\ae{}rgaard P. 1993, \apj,
411, 34
\reference J\o{}rgensen I., Franx M., \& Kj\ae{}rgaard P. 1996,
\mnras, 280, 167
\reference Kelson, D.D., \etal\ 1999a, \apj, submitted
\reference Kelson, D.D., \etal\ 1999b, \apj, in press
\reference Kelson, D.D., \etal\ 1999c, \apj, in press
\reference Kelson, D.D. 1999, in preparation
\reference Tinsley, B.M. \& Gunn, J.E. 1976, ApJ, 203, 52
\reference van Dokkum, P.G., \etal\ 1998, \apj, 500, 714
\reference Vazdekis, A., \etal\ 1996, \apjs, 107, 306
\reference Worthey, G. \& Ottavianni, D.L. 1997, \apjs, 111, 377
\end{references}
\end{document}